\documentclass[conference]{acmsiggraph}
\usepackage[numbered]{algorithm}
\usepackage[noend]{algorithmic}
\usepackage[super]{nth}
\usepackage{wrapfig}
\usepackage{float}
\usepackage{fixltx2e}
\usepackage{bbm}
\usepackage{siunitx}
\usepackage{tikz}

\usetikzlibrary{positioning}
\usetikzlibrary{calc}

\def\BibTeX{{\rm B\kern-.05em{\sc i\kern-.025em b}\kern-.08em
    T\kern-.1667em\lower.7ex\hbox{E}\kern-.125emX}}

\TOGonlineid{0358}


\TOGvolume{0}
\TOGnumber{0}

\title{Towards a Drone Cinematographer: \linebreak Guiding Quadrotor Cameras using Visual Composition Principles}

\author{Niels Joubert$^{*}$\\Stanford University \and Jane L. E\thanks{Niels Joubert and Jane E contributed equally to this work.}\\Stanford University \and Dan B Goldman\\Google \and Floraine Berthouzoz\\Adobe Research \and Mike Roberts\\Stanford University \and James A. Landay\\Stanford University \and Pat Hanrahan\\Stanford University}
\pdfauthor{Jane L. E, Niels Joubert, Dan B Goldman, Floraine Berthouzoz, Mike Roberts, James A. Landay, Pat Hanrahan}


\floatstyle{boxed}
\newfloat{Listing}{t}{ext}

\setcopyright{rightsretained}

\copyrightyear{2016}

\conferenceinfo{SIGGRAPH Asia 2016 }{} 
\isbn{978-1-4503-ABCD-E/16/07} 
\doi{http://doi.acm.org/10.1145/9999997.9999999}

\begin{document}

\definecolor{purple}{RGB}{123,102,210}

\newcommand{\niels}[1]{\textbf{\textcolor{magenta}{NJ: #1}}}
\newcommand{\pat}[1]{\textbf{\textcolor{blue}{PH: #1}}}
\newcommand{\jane}[1]{\textbf{\textcolor{purple}{JE: #1}}}
\newcommand{\dan}[1]{\textbf{\textcolor{green}{DG: #1}}}
\newcommand{\mike}[1]{\textbf{\textcolor{cyan}{MR: #1}}}

\setlength{\textfloatsep}{15pt}
\setlength{\dbltextfloatsep}{5pt}
\setlength{\intextsep}{5pt}
\setlength{\abovecaptionskip}{5pt}
\setlength{\belowcaptionskip}{10pt}
\setlength{\parskip}{5pt}


 \teaser{
   \includegraphics[width=1\linewidth]{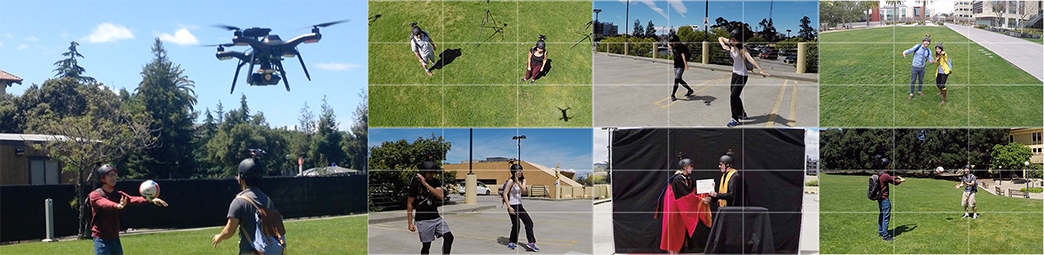}
   \caption{We present an end-to-end system for capturing well-composed footage of two subjects with a quadrotor in the outdoors. On the left, we show the quadrotor filming two subjects. To the right, are static shots captured by our system, covering a variety of perspectives and distances. We demonstrate people using our system to film a range of activites--pictured here: taking a selfie, playing catch, receiving a diploma, and performing a dance routine.
   \label{fig:teaser}}
 }


\maketitle

\begin{abstract}

We present a system to capture video footage of human subjects in the real world.
Our system leverages a quadrotor camera to automatically capture well-composed video of two subjects.
Subjects are tracked in a large-scale outdoor environment using RTK GPS and IMU sensors. 
Then, given the tracked state of our subjects, our system automatically computes static shots based on well-established visual composition principles and canonical shots from cinematography literature. 
To transition between these static shots, we calculate feasible, safe, and visually pleasing transitions using a novel real-time trajectory planning algorithm. 
We evaluate the performance of our tracking system, and experimentally show that RTK GPS significantly outperforms conventional GPS in capturing a variety of canonical shots. 
Lastly, we demonstrate our system guiding a consumer quadrotor camera autonomously capturing footage of two subjects in a variety of use cases. 
This is the first end-to-end system that enables people to leverage the mobility of quadrotors, as well as the knowledge of expert filmmakers, to autonomously capture high-quality footage of people in the real world.

\end{abstract}

\pagebreak
\section{Introduction} \label{sec_introduction}

Quadrotors are enabling new forms of cinematography. 
Small unmanned aerial vehicles can fly to unique vantage points,
and their maneuverability allow them to fly along acrobatic trajectories.
Moreover, quadrotors with high quality cameras are relatively inexpensive,
making them accessible to serious amateurs.
As evidence of their popularity,
there are now film festivals dedicated exclusively to films shot with quadrotors.

In this paper, we investigate the use of quadrotors
to film people doing everyday activities, 
including sports and dance.
Our basic idea is to create a semi-autonomous quadrotor
camera system that positions itself relative to
the people in a scene according to rules of cinematography.
This allows the quadrotor to capture well-composed footage of people
without needing another person to manually fly the quadrotor.
Flying a quadrotor is challenging and requires skill,
and needing a pilot requires an additional person.
Our work builds on the capabilities of recent commercial
systems that have a \emph{follow-me} mode.
We seek to enable a more sophisticated \emph{cinema-mode},
where the quadrotor seeks to capture visually pleasing
shots of the activity being undertaken.

We focus on filming scenes with one or two people, where both people are fairly stationary. Under these simplifying assumptions, we demonstrate the efficacy of our idea for a specific set of scenarios, such as two people taking a ``selfie'', playing catch, or performing a hip-hop dance routine. 

We draw upon cinematographic practice, 
and past work in computer graphics
that adds visual composition principles to virtual camera controllers. 
Cinematographers have devised a small set
of canonical shot types (e.g., apex, internal, and external),
and computer graphics researchers have developed algorithms
for placing the camera to generate these shot types.
The result is that if we know the locations of the subjects,
we can compose visually pleasing footage
by correctly placing the camera.

However, there are several challenges 
to using visual composition principles and canonical shots with quadrotors, all related to the fact that the quadrotor
is a physical device moving in the real world:
A quadrotor must obey the laws of physics, which 
constrain how it can fly.
Moreover, the quadrotor must know where the subjects are to be able to film them.
Most importantly, the quadrotor must not fly into 
people and cause them harm.

We overcome these challenges 
through accurate tracking and a unique trajectory planning algorithm.
We track the position of the quadrotor and two people
using Real Time Kinematic (RTK) GPS.
RTK GPS allows us to estimate positions to within 2\,cm accuracy, 
which is much better than the roughly 2\,m accuracy of conventional GPS.
Given the sizes and distances involved, this increased accuracy is essential to 
place the subjects correctly within a frame.

Our trajectory planning algorithm builds upon previous work in
designing and optimizing quadrotor camera trajectories.
Like previous work,
we require that the trajectory flown by the quadrotor
obey the laws of physics. 
This requires that the path be $C^4$ continuous,
and that the movement along the path not exceed a maximum velocity.
The main new technical contribution in this paper 
is a method to move between camera shots safely;
that is, we guarantee 
that the position of the quadrotor relative to a subject
is greater than a minimal distance.
We enforce this no-fly ``safety sphere"
while maintaining a pleasing composition of the image.

We present the first end-to-end system that leverages composition principles and canonical shots to guide autonomous quadrotor cameras filming people in the real world.

\section{Related Work} \label{sec_relwork}

\paragraph{Autonomous Cinematography in Virtual Environments}
Automatically capturing visually pleasing footage in a virtual environment is a classic problem in computer graphics~\cite{CON08}. A common approach is to find camera poses based on principles from cinematography literature~\cite{He96,courty2003cinematography,li2005interactive}.

He et al.~\shortcite{He96} present a set of heuristics to pose virtual cameras based on visual composition principles, and use these heuristics to design \textsc{The Virtual Cinematographer}. 
We extend their method to also consider safety of subjects, while maintaining the same visual composition principles. 
 
Interpolating between multiple camera poses is a well-studied subproblem of autonomous cinematography. Recently, Lino and Christie~\shortcite{Lino15} demonstrated an analytic method for interpolating between viewpoints of subjects in a way that produces visually pleasing results. Their main insight was to define a visual interpolation space relative to each subject, and analytically compute a resulting camera path. They show how to solve for a camera position given two screen-space positions and a distance to the closest subject, also known as the Blinn spacecraft problem~\cite{blinn1988looking}. Lino and Christie's approach has been used to generate smooth trajectories using an iterative approximation approach~\cite{Galvane:2015:CAC:2822013.2822025}, and as a target for the Prose Storyboard Language~\cite{galvane:hal-01067016}. Their approach has also been extended to force-based camera models with soft constraints~\cite{galvane2013steering}. We build on Lino and Christie's visual interpolation space to design a new method that also produces visually pleasing results, while we specifically respect the safety of \textit{both} subjects and the requirements of quadrotor hardware. 

\paragraph{Autonomous ``Follow-me'' Quadrotors} There is significant commercial and research interest in designing quadrotors to autonomously capture video of subjects. Naseer et al.~\shortcite{naseer2013followme} demonstrates a quadrotor that follows a person using RGB-D depth tracking. The \textsc{3DR Solo}, \textsc{DJI Phantom}, \textsc{Yuneec Typhoon}, \textsc{AirDog}, and \textsc{Ghost Drone} all feature a ``follow-me'' mode that tracks subjects either visually or using GPS. Existing approaches only attempt to keep the subject visible, and rely on the operator to pose the camera relative to the subject. In contrast, our approach incorporates high-level cinematography principles to \emph{automatically} find visually pleasing poses.

Tracking subjects in the context of controlling a quadrotor is an area of active study. A promising approach uses vision or depth sensors placed on the quadrotor to track subjects~\cite{teuliere2011chasing,naseer2013followme,lim2015monocular,Coaguila2016}. An alternative approach is place sensors on subjects, relieving the quadrotor from maintaining a visual line of sight to all subjects. Inspired by work using centimeter-accurate RTK GPS to study human movement~\cite{terrier2005useful}, we use RTK GPS combined with Inertial Measurement Unit (IMU) sensors to track subjects and demonstrate its efficacy for automating cinematography.

Coaguila et al.~\shortcite{Coaguila2016} considered the problem of moving a quadrotor camera to capture well-composed video of a subject using visual tracking of the face. Whereas their work is specific to capturing a full-frontal composition of a single moving subject, our system is concerned with capturing a variety of cinematic shots of two subjects.

\begin{figure*}[th!]
\centering
\includegraphics[width=1\linewidth]{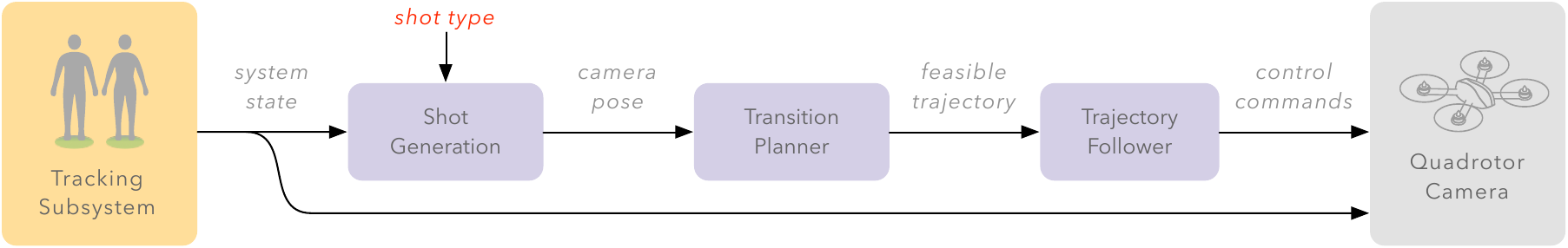}
\caption{
Major technical components of our real-time autonomous cinematography system: Our tracking subsystem estimates poses of subjects and the quadrotor camera in the real world. Whenever a new shot type is provided by a user (shown in red), our system generates a camera pose that satisfies visual composition principles and physical placement constraints. To move the camera from its current pose to this new camera pose, a feasible, safe, and visually pleasing transition is planned. Finally, a sequence of quadrotor and gimbal commands control the quadrotor camera autonomously. 
\label{fig:sysdiagram}}
\end{figure*}

\paragraph{Trajectory Planning Methods for Quadrotor Cinematography}
Recent work in the graphics community investigates methods to control quadrotor cameras. Joubert et al.~\shortcite{Joubert15} introduce a tool for interactively designing quadrotor camera trajectories. Gebhardt et al.~\shortcite{Gebhardt:2016} demonstrate a method for generating trajectories according to high-level user goals. Roberts and Hanrahan~\shortcite{roberts:2016} demonstrate a method for generating feasible trajectories from infeasible inputs. These methods remove the need for manually piloting a quadrotor to capture cinematography. However, these tools rely on the user to specify keyframes before capture, and do not allow the user to use people's positions in their shot reference frames.

A common approach for planning quadrotor trajectories is to generate $C^4$ continuous splines~\cite{deits2015efficient,Joubert15,richter:2013,mellinger:2011}. 
Our system relies on the same property, using the quadrotor camera model introduced by Joubert et al.~\shortcite{Joubert15}.

Several of these methods also plan quadrotor trajectories around obstacles~\cite{Gebhardt:2016,deits2015efficient,richter:2013,mellinger:2011}, including state-of-the-art methods that are fast enough to run in real time~\cite{allen2016real}. Our approach to trajectory planning is complementary to this work, since our system plans trajectories around two spherical obstacles representing our subjects, while also considering visual aesthetics along the trajectory. Our approach is also fast enough to run in real time.

Galvane et al.~\shortcite{wiced.20161097} presents a system that controls a quadrotor camera to capture one or two subjects. Similar to our system, Galvane encodes cinematic properties to intelligently frame subjects, and relies on Lino and Christie's method for finding camera poses given screen-space constraints. Unlike our method, Galvane does not guarantee a minimum distance to both subjects, does not guarantee feasibility of their resulting trajectories, and depends on an indoor tracking system.

\paragraph{Safer Quadrotors}
Several companies are developing safer quadrotor hardware, such as the \textsc{Parrot AR}, \textsc{Hover Camera} and \textsc{Flyability Gimball}. These quadrotors reduce the potential harm of a collision by enclosing the propellers inside a safety mesh or shell. Recently, the first consumer quadrotors with active obstacle avoidance became available. The \textsc{DJI Phantom 4} and \textsc{Yuneec H} both attempts to detect an obstacle and take evasive action. However, these systems do not attempt to produce visually pleasing cinematography while avoiding obstacles. 

\paragraph{Autonomous Cinematography using Robotic Cameras}
More broadly, guiding robotic cameras using visual composition principles has been investigated in the robotics literature. Some of this work also crops the resulting footage to improve visual composition when the robot cannot place itself in the desired pose~\cite{campbell2005leveraging}. However, this work mostly focuses on wheeled or stationary robots~\cite{byers2003autonomous,ahn2006robot,kim2010automatically,gadde2011aesthetic}. In contrast, our system explicitly considers the dynamics of quadrotors when planning shots.

\section{Design Goals and Challenges} \label{sec_design}

We design our system to achieve the following visual composition goals:

\paragraph{Employ Canonical Shots} The literature of cinematography offers numerous high-level composition
principles that guide the framing of a set of commonly used shots~\cite{Arijon76,He96,Rubin09}. These shots
specify where subjects lie within the frame, and implicitly define the relative placement of
the camera with respect to the subjects. Following this approach, we implement a set of canonical static shots. These shots place the camera at a fixed pose, allowing subjects some freedom to move in the frame. We also take care to respect the compositional principle of the \emph{rule of thirds}: the focal point of a shot is placed at the intersection of horizontal and vertical lines splitting the screen into thirds.

\paragraph{Maintain Compositional Continuity} 
Moving shots and transitions should be preplanned such that start and end frames are compositionally balanced, and intermediate framings should vary smoothly, with subjects moving in roughly straight lines from one camera movement to the next. We wish to avoid indecisive and jerky motions: 
Movement that is too fast can be hard on viewers' eyes, and can detract from the content of the frames. In addition, we seek to preserve the \emph{line of action} between the two subjects:
Throughout a shot sequence, the camera stays on the same side of this line to encourage visual continuity. 
The rationale for this principle is that if the camera switches sides, the subjects will switch left and right sides in the frame, which is disorienting for the viewer.

To achieve our stated design goals, our system must overcome the following technical challenges:

\paragraph{Construct and Maintain a Virtual Representation of the Scene} Our virtual representation of the scene must be accurate enough to plan shots with the intended visual composition. This implies that the system must accurately track the pose of both subjects and camera. Our system supports capturing shots containing one or both people, and therefore our tracking system cannot assume both subjects are always visible to the primary camera.

\paragraph{Plan Safe Camera Locations} Based on the virtual scene, our system will plan locations for the camera in the physical world. In consideration of both safety and personal space, we introduce a safety constraint. This constraint states that we only choose camera positions outside of exclusion zones where the camera must not be placed. We call these zones, centered on each subject, \emph{safety spheres}, represented as a minimum distance constraint in 3D space. 

\paragraph{Plan Visually Pleasing Transitions} In contrast to a virtual camera, a physical camera cannot be immediately placed at a new location: It has to transition in space between poses. We implement transitions between shot locations, planned such that they attempt to maintain pleasing composition throughout. During these transitions, the quadrotor camera must also maintain the safety minimum distance constraint to both subjects.

\paragraph{Place the Camera According to Plan} Finally, the virtual plan of static shot locations and transitions must be executed by a physical quadrotor control system in real time.
\section{Technical Overview} \label{sec_overview}


We provide an overview of the major technical components of our system in Figure~\ref{fig:sysdiagram}. At the core of our system is a shot generator that produces well-composed static shots of two subjects, described in Section~\ref{sec_shots}. We build this shot generator based on a set of canonical shot types and visual composition principles. This shot generator enables users to specify a desired camera pose at a high level, using terminology from the cinematography literature. Given a canonical shot type, our shot generator produces a static camera pose consisting of a look-from and look-at point, taking care to ensure the resulting pose is safe with respect to both subjects. Our system then places and holds a camera at this pose, recording video. 

We prototype a simple user interface driven by a visualization of the virtual scene representation and a simulation of our robotic camera. Our interface displays a 3D rendering of the current virtual representation of the scene from the perspective of the quadrotor camera. A user can select any shot type, and virtually see the resulting shot. The user can also issue shot types to the real quadrotor camera. Our system will then place the virtual camera at a new static camera pose corresponding to the selected shot type.

To place the camera at a new static camera pose, our system creates a transition from the current camera pose to this new pose. This transition needs to take into account the visual contents of video recorded during a transition, respect the safety of subjects, and adhere to the capabilities of quadrotors. With this in mind, we design an algorithm for synthesizing quadrotor camera trajectories between two static camera poses (Section~\ref{sec_transitions}). At a high level, our approach is to optimize a \emph{blend} of easy-to-generate basis trajectories by solving a constrained nonconvex optimization problem. Using this algorithm, we produce a look-at and look-from trajectory for our quadrotor camera.

Our shot generator produces a camera pose relative to subjects, and thus needs to know the pose of each subject. As discussed in Section~\ref{sec_tracking}, our system tracks subjects by having them wear a helmet containing high-accuracy RTK GPS and inertial measurement unit (IMU) sensors. We use the same tracking system to accurately localize our quadrotor camera.

Lastly, our system captures shots by issuing control commands to a quadrotor camera. These control commands takes the form of look-from and look-at setpoints driving a feedback controller running on a real-world quadrotor. During a static shot, our system holds a quadrotor camera at a fixed look-from and look-at setpoint until the user commands a new shot. During a transition to a new shot, our system sends a stream of look-from and look-at samples along the transition trajectory, moving the quadrotor camera.

Throughout our system, we consider various approaches to keep our subjects safe in the presence of a quadrotor aircraft. When our system places static shots, it keeps the quadrotor a safe distance from our subjects. While a camera is at a static camera pose, subjects are free to move around and the camera will not change its position. When our system plans a transition, it ensures the resulting trajectory stays a safe distance from subjects, but assumes the subjects will not leave their safety spheres during a transition. Overall, our system does not prevent a subject from intentionally colliding with the quadrotor. We expect advances in dynamic obstacle avoidance to address this problem. For the purposes of this paper, we feel it is reasonable to assume a benevolent subject that is willing to remain fairly stationary within the bounds of the safety sphere.








\section{Modeling Subjects and Cameras}

\begin{figure}[t]
  \centering
  \includegraphics[width=2.6in]{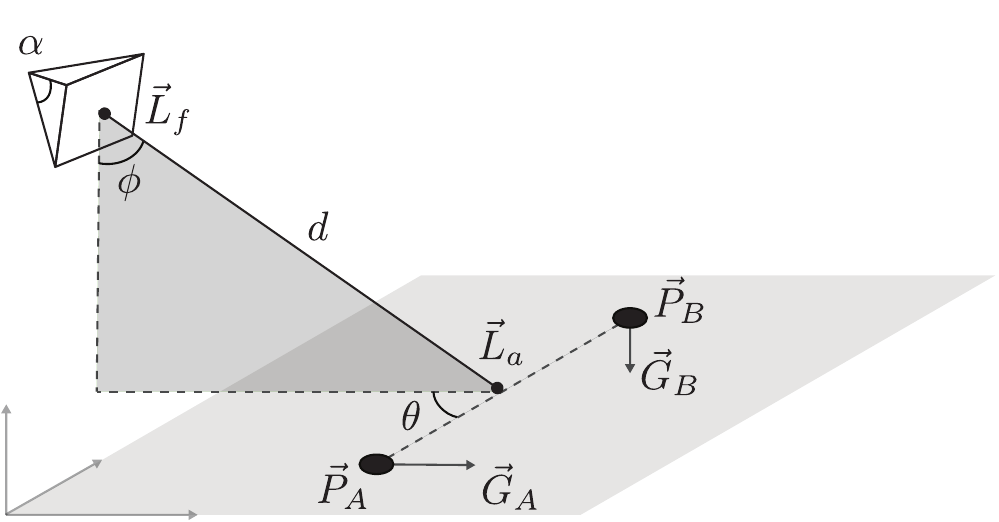}
  \caption{
Overview of our camera and subject model.
Each subject has a position and a gaze vector,
$\vec{P}$ and $\vec{G}$.
We model our camera as a look-from point $\vec{L}_f$,
a look-at point $\vec{L}_a$, and a field of view $\alpha$.
We also introduce angles $\theta$ and $\phi$ to describe
angles between the direction that the camera is pointed
and the line of action,
and $d$ to indicate the distance between the look-at and look-from points.
}
  \label{fig:model}
\end{figure}

In this section, we introduce the subject and quadrotor camera models used in our system, shown in Figure~\ref{fig:model}.

\subsection{Subject Model} 

Each subject is modeled 
as a position $\vec{P}_i$ 
and gaze vector $\vec{G}_i$ 
that represents the subject's head position and orientation.
Our tracking system, presented in Section~\ref{sec_tracking},
estimates these positions and orientations.
We calibrate our tracking system 
so that the position of the subject corresponds 
to the center of their head at eye level. 
Each subject is further described by their height, and by a minimum distance value $d_{min}$ defining the subject's safety sphere.
Our system will not place the quadrotor camera 
closer than $d_{min}$ to a subject.

\subsection{Quadrotor Camera Model} 

We use the joint quadrotor and camera model 
introduced by Joubert et al.~\shortcite{Joubert15},
which models a camera on a gimbal attached to a quadrotor aircraft.
This model has two important implications for us.
First, this model enables our system 
to specify the behavior of a quadrotor camera 
using \emph{look-from} and \emph{look-at} world space points
($L_f$ and $L_a$ in Figure~\ref{fig:model}).
We fix the $up$ vector to be vertical.

To model moving quadrotors and subjects,
we allow the look-at and look-from points
to follow a trajectory.
To respect the dynamics and physical limits of the quadrotor,
these look-at and look-from trajectories must be $C^4$ continuous,
and the velocity along the look-from trajectory 
must be less than the maximum speed at which the quadrotor can fly.

Finally, our physical camera model has a fixed field of view $\alpha_\text{max}$.


\section{Generating Static Shots} \label{sec_shots}

\begin{figure*}[th!]
\centering
\includegraphics[width=1\linewidth]{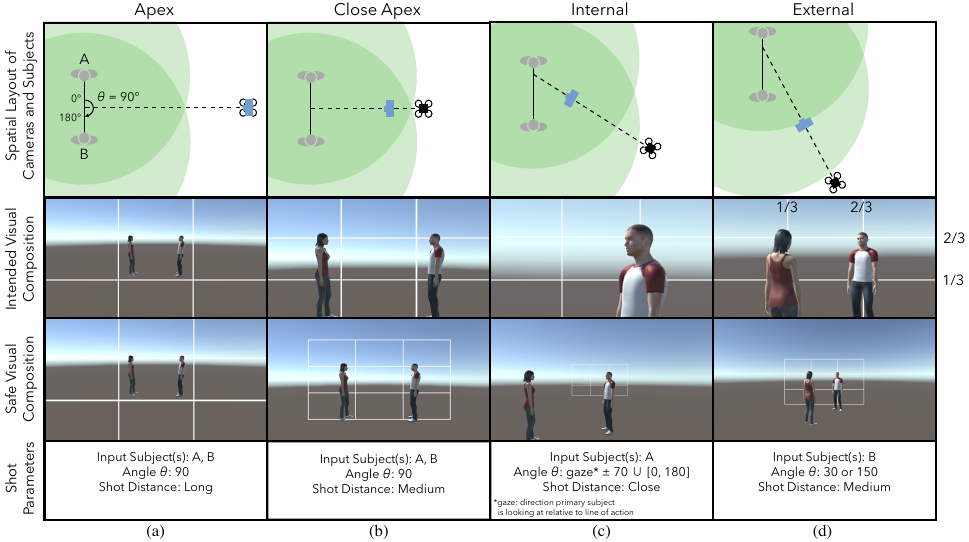}
\caption{
This table shows the four main types of shots 
we implemented in our system. 
The top row shows the spatial layout of each shot from a bird's eye view.
The blue camera is the goal virtual camera position,
and the black quadrotor shows the same shot from a safe distance. 
The second row shows the intended visual composition, applying the rule of thirds.
In the third row we show the same shot after applying our minimum distance constraint.
We move the camera to a safe distance 
by decreasing the field of view,
while maintaining the size of the subjects by cropping the frame.
Finally, we list the parameters that define each shot.
In this illustration, the pitch angle $\phi = 0$.
\label{fig:staticshotsoverview}}
\end{figure*}

Our system is designed to capture canonical shots 
which adhere to principles gleaned from the cinematography literature.
Here we describe the set of shots we selected,
and how they are implemented in our system.


\subsection{Defining Shots}

The inputs to the shot selection system are the positions and gaze directions
of the two subjects, and the desired type of shot.
The outputs of the shot selection system
are the look-from and look-at points, and the field of view of the camera.

A shot type is defined by the input subject(s), shot distance, and orientation angles.
Here we describe these parameters and how they impact the outputs for a given shot.

If the shot has a single \emph{input subject} 
(e.g. the internal and external shots described below),
we place our camera so that this primary input subject
is at a screen space position 
that follows the compositional principle of the \emph{rule of thirds}.
More specifically, the eyes of the subject 
are placed at the intersection of horizontal and vertical lines 
splitting the screen into thirds.
Subjects facing to screen right lie along the left vertical one-third line,
and subjects facing to screen left lie along the right vertical one-third line.
If a shot has multiple input subjects (e.g. an apex shot),
we frame both subjects by placing the camera 
so the average position of the eyes of the two subjects 
lies on the center of the upper horizontal two-thirds line.

We next set the distance of the camera to the primary subject,
or, in the case of two subjects, 
to the average position.
We specify \emph{shot distances} qualitatively 
using well-defined cinematographic conventions:
close, medium, or long~\cite{Arijon76}.
These distances are defined by the portion of a subject's body 
that should appear in frame---specifically,
we represent these as the approximate number of heads
below the horizon line
(close is 2.5, medium is 4, and long is 7.5).
We geometrically calculate an absolute distance $d$ based on shot distance and the subject(s)' average height.

The orientation angle $\theta$ defines the yaw of the camera 
relative to the line of action.
Cameras are placed on the same side of the line of action.
The system initially chooses the side that sees more of the subject(s)' faces,
which can be determined from the gaze directions,
and keeps the camera on this side.
We also have an angle $\phi$ that defines the pitch of the camera. 
This is usually set to 0 degrees,
generating shots with a straight-on view of subjects.

The look-from and look-at point for a shot is calculated 
from the screen space position of the subjects,
the distance of the camera, and
the orientation angle.
An approximate look-at point is placed on the line of action,
and an approximate look-from point is placed relative 
to the approximate look-at point using $\theta$, $\phi$, and $d$.
This places the camera at the correct orientation and distance from the subject,
but does not yet guarantee the subject appears 
in the correct screen space position.
We shift the approximate look-from and look-at point 
to move the subjects to the correct screen space position in the frame.



The distance of the camera to the subject depends 
on the field of view of the camera
and the desired size of the subject in the frame.
Unfortunately, this may cause the camera
to be placed inside the safety spheres surrounding the subjects.
We fix this hazard by moving the camera further away 
until it is outside the safety spheres.
Moving the camera further away causes the subject size to shrink,
and the composition to change.
To compensate for this change,
we calculate a crop, $\alpha \leq \alpha_\text{max}$,
that maintains the visual composition of the shot. It is possible for the crop to be so extreme that the resolution loss makes the resulting footage practically unusable even though subjects are correctly framed.  Fortunately, our system can use the same approach for quadrotors with optical zoom lenses to change the view of view without incurring resolution loss.

\subsection{Types of Canonical Shots}

We chose to implement four main shots in our system:
apex, close apex, internal, and external.
These shots are adapted from the camera modules in He et al.~\shortcite{He96}. 
Figure~\ref{fig:staticshotsoverview} shows these four shot types,
the relative spatial placement of the camera and subjects,
as well as the resulting visual compositions.

\begin{itemize}

\item{\textbf{Apex}} A long shot of both subjects,
vertically centering characters in the frame.
The subjects' average eye level is placed centered horizontally at the two-thirds line (Figure~\ref{fig:staticshotsoverview} (a)).

\item{\textbf{Close apex}} A medium shot of both subjects,
framed similarly to the Apex shot (Figure~\ref{fig:staticshotsoverview} (b)).

\item{\textbf{Internal}} A close shot of a single input subject,
oriented relative to gaze to guarantee a semi-frontal view of the primary subject.
The subject is placed on one of the vertical thirds lines such that the majority of empty screen space is in front of her (Figure~\ref{fig:staticshotsoverview} (c)).

\item{\textbf{External}} A medium shot of a single input subject,
looking over the shoulder of the other subject.
If the primary subject is on the left side,
she is placed at the one-thirds line 
with the other subject in the right third of the frame,
and vice versa (Figure~\ref{fig:staticshotsoverview} (d)).

\item{\textbf{Apex From Above, External From Above}} We also 
implemented alternate versions of the Apex and External shots,
but placed above subjects to mimic canonical top-down shots~\cite{Rubin09}.
These are implemented with the same parameters as the shot types described above, but with
the pitch angle, $\phi$, set to place the camera looking down from above the subjects.

\end{itemize}

\section{Transitioning Between Shots} \label{sec_transitions}



%


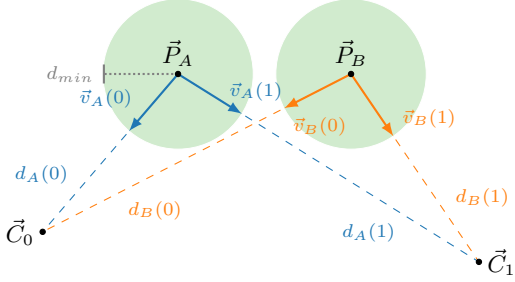
\begin{figure}[th!]
\centering
\begin{tikzpicture}


\coordinate (PA) at (1,0);
\coordinate (PB) at (3.3,0);
\coordinate (CA) at (-0.8,-2.1);
\coordinate (CB) at (5,-2.5);
\coordinate (PAs) at (0,0);

\definecolor{safetygreen}{cmyk}{0.62,0,0.86,0}

\definecolor{tblue}{RGB}{31,119,180}
\definecolor{torange}{RGB}{255,127,14}

\filldraw[color=green!0, fill=safetygreen, opacity=0.3, line width=0mm](PA) circle (1);
\filldraw[color=green!0, fill=safetygreen, opacity=0.3, , line width=0mm](PB) circle (1);

\draw[densely dotted, gray, thick, -|] (PA) -- (PAs) node[anchor=east] {\small $d_{min}$};

\draw[thin, dashed,tblue] (PA) -- (CA) node[near end, above left] {\small $d_A(0)$};
\draw[thin, dashed,tblue] (PA) -- (CB) node[near end, below left ] {\small $d_A(1)$};
\draw[thin, dashed,torange] (PB) -- (CA) node[near end, below right] {\small $d_B(0)$};
\draw[thin, dashed,torange] (PB) -- (CB) node[near end, above right] {\small $d_B(1)$};

\draw[thick, -latex,tblue]      (PA) -- ($ (PA) !1cm! (CA) $) node[near end, above left] {\small $\vec{v}_A(0)$};
\draw[thick, -latex,tblue]      (PA) -- ($ (PA) !1cm! (CB) $) node[near end, above right=-0.1cm] {\small $\vec{v}_A(1)$};
\draw[thick, -latex,torange] (PB) -- ($ (PB) !1cm! (CA) $) node[below right] {\small $\vec{v}_B(0)$};
\draw[thick, -latex,torange] (PB) -- ($ (PB) !1cm! (CB) $) node[above right] {\small $\vec{v}_B(1)$};

\filldraw[black] (PA) circle (1pt) node[anchor=south] {$\vec{P}_A$};
\filldraw[black] (PB) circle (1pt) node[anchor=south] {$\vec{P}_B$};
\filldraw[black] (CA) circle (1pt) node[anchor=east] {$\vec{C}_0$};
\filldraw[black] (CB) circle (1pt) node[anchor=west] {$\vec{C}_1$};


\end{tikzpicture}
\caption{Here we show the terms we use to construct basis vector paths. We assume we are given two subject positions $\vec{P}_A$, $\vec{P}_B$ and an initial and final camera position $\vec{C}_0$, $\vec{C}_1$. In blue, we show the terms that generate a basis path for Subject A. We extract an initial and final vantage vector $\vec{v}_A(0)$, $\vec{v}_A(1)$, and an initial and final distance $d_A(0)$, $d_A(1)$. We linearly interpolate from $d_A(0)$ to $d_A(1)$, and spherical linearly interpolate from $\vec{v}_A(0)$ to $\vec{v}_A(1)$. Scaling the interpolated vantage vector by the interpolated distance as we interpolate produces a basis path. In orange, we show the same quantities relative to Subject B. \label{fig:sphericaltrajectories}}
\end{figure}

\begin{figure}
%
%
\centering
\includegraphics[width=3.3in]{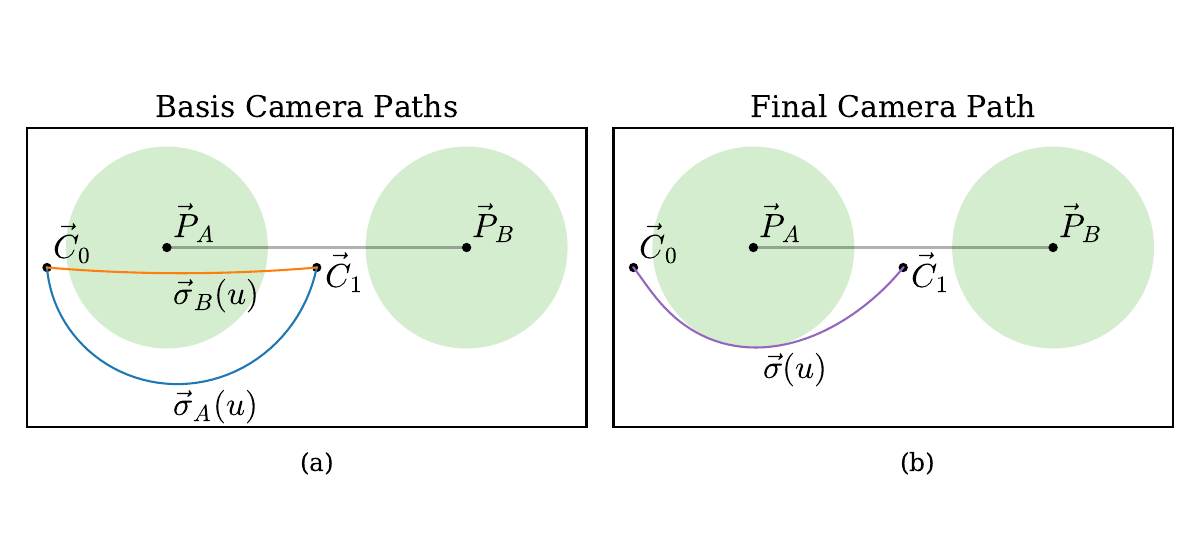}
\caption{Blending between quadrotor camera trajectories. 
(a) A top-down view of our basis camera paths, generated using spherical interpolation 
around Subject A (blue path) and Subject B (orange path), respectively.
The green circles represent the safety sphere of each subject.
Note that the orange basis path violates the minimum distance constraint (green circle) around Subject A.
(c) We find a final camera path by blending these two basis paths,  enforcing the constraint that the final path is outside \emph{both} unsafe regions.
\label{fig:optimizationsteps}}
\end{figure}

In this section, we consider the problem of moving a quadrotor camera from
one static shot to another.
Specifically, we want to find a quadrotor camera trajectory 
that maintains a visually-pleasing composition, 
respects the dynamics and physical limits of our hardware,
and ensures safety of our subjects.
Our main insight is to avoid solving a general trajectory optimization problem in the full state space of the quadrotor.
Instead, we \emph{blend} between two easy-to-generate and visually-pleasing \emph{basis} trajectories.
This approach is more computationally efficient than solving a general trajectory optimization problem, and produces safe and visually pleasing trajectories.

We summarize our method for transitioning between static shots as follows.
First, we generate a pair of basis paths by adapting a composition-aware interpolation technique introduced by Lino and Christie~\shortcite{Lino15}.
These basis paths produce visually pleasing results, but might get too close to the subjects.
We produce a final path that respects our minimum distance constraints, by optimizing a \emph{blending function} that blends between our basis paths.
We then apply an easing curve to our final path, producing a final look-from trajectory.
We generate a look-at path by linearly interpolating look-at points in world space.
We apply the same easing curve (as was applied to our look-from path) to our look-at path to produce a final look-at trajectory.

We assume we are given as input a start and end camera position $\vec{C}_{0}$ and $\vec{C}_{1}$, as well as start and end look-at points.
We assume the start and end positions of the look-at point do not change during the transition.
We also assume the start and end camera positions are safe -- that is, the distance from the start and end camera position to each subject is greater than $d_{min}$.


\subsection{Generating Basis Paths}



We assume that the paths we consider in this section are parameterized by a scalar \emph{path parameter} $u$.
We define two basis paths, one for each subject.
We define the camera-to-subject distance at each point along each basis path as $d_i(u)$, where $i$ is an index that refers to each subject.
We set $d_i(u)$ to be the linear interpolation between the camera-to-subject distances of our initial and final camera positions $\vec{C}_0$ and $\vec{C}_1$.
Likewise, we define the vantage-vector at each point along the basis path as $\vec{v}_i(u)$.
We set $\vec{v}_i(u)$ to be the spherical linear interpolation of the normalized camera-to-subject vantage vectors corresponding to our initial and final camera positions.
We define our two basis paths $\vec{\sigma}_i(u)$ as follows.
\begin{equation}
        \vec{\sigma}_i(u) = \vec{P}_i + d_i(u)\cdot\vec{v}_i(u)\quad \text{for}\ u \in [0,1], i = \{A,B\}
\end{equation}
This construction is shown in Figure~\ref{fig:sphericaltrajectories}.

Our basis paths have the following useful properties:

\begin{itemize}
\item $\vec{\sigma}_i(u)$ is $C^\infty$ continuous with respect to $u$. This is because spherical linear interpolation between two vectors and linear interpolation between two points are both $C^\infty$ continuous interpolation schemes.
This property is useful, since trajectories must be at least $C^4$ continuous with respect to time in order to satisfy the quadrotor dynamics.

\item It is guaranteed that $\vec{\sigma}_A(u)$ will never get too close to subject $A$, and $\vec{\sigma}_B(u)$ will never get too close to subject $B$.
This is because our start and end camera positions satisfy the minimum distance constraint, and we linearly interpolate distance.
This property is useful, because it suggests that we can generate a path that never gets too close to \emph{either} subject by blending between our basis paths.
\end{itemize}

\subsection{Optimal Blending of Basis Paths}


In the previous section we generated two paths, one relative to each subject.  Previous work averages these two paths together to produce a final path.  Unfortunately, the resulting path can violate our minimum distance constraint (see Figure 6).

We introduce a blend function $w(u)$
that blends the two basis paths into a final path $\vec{\sigma}(u)$ as follows,

\begin{equation} \label{eq:blendedtrajectory}
        \vec{\sigma}(u) = w(u) \cdot \vec{\sigma}_A(u) \; + \; (1 - w(u)) \cdot \vec{\sigma}_B(u)
\end{equation}

We now use constrained optimization to find a good blend function.
We seek a blend between the two basis paths
(1) that is as close as possible to the two input paths,
and (2) obeys the minimum distance constraint.
During this optimization procedure,
we also enforce $C^4$ continuity (and hence, $C^4$ continuity of our final path),
and we optimize the overall smoothness of our blend.






\paragraph{Enforcing $C^4$ Continuity}

In order to enforce $C^4$ continuity,
we discretize our blend function $w(u)$
into a sequence of $n$ sample points $w_k$ 
where $k = 1 \ldots n$ (thus $u = \frac{k}{n}$).

Following the approach outlined by 
Roberts and Hanrahan~\shortcite{roberts:2016},
let $\mathbf{w}_k$ be the value and 
the first four derivatives of $w(u)$ at each sample point $k$.
Let $v_k$ be the 5\textsuperscript{th} derivative 
of $w(u)$ at sample point $k$.
Let $du$ be the delta between successive sample points.
We enforce $C^4$ continuity of our blend as follows,

\begin{equation*}
\mathbf{w}_{k+1} = \mathbf{w}_{k} + (\mathbf{M}\mathbf{w}_{k} + \mathbf{N}v_{k})du
\end{equation*}
\begin{equation} \label{eqn:continuity}
\text{where~~~~}
\mathbf{M} = \begin{bmatrix}
0&1&0&0&0\\
0&0&1&0&0\\
0&0&0&1&0\\
0&0&0&0&1\\
0&0&0&0&0\\
\end{bmatrix}
~~~~
\mathbf{N} = \begin{bmatrix}
0\\
0\\
0\\
0\\
1\\
\end{bmatrix}
\end{equation}
\begin{equation*}
\text{subject to~~~~}v^{\text{min}} \leq v_k \leq v^{\text{max}}
\end{equation*}

Similarly to Roberts and Hanrahan~\shortcite{roberts:2016},
we introduce $v^{\text{min}}$ and $v^{\text{max}}$ 
to control how much $\frac{d^4 w}{du^4}$ 
is allowed to vary between sample points while still considered continuous.
In our implementation, we heuristically set these values inversely proportional to the number of samples $n$ of our blend.

\paragraph{Optimization Problem}


Stating our optimization problem formally,
let $\mathbf{W}$ be the concatenated vector of 
decision variables $\mathbf{w}_k$ and $v_k$ 
across all sample points $k = 1, \ldots, n$. Let $\lambda$ be a parameter that trades off between smoothness and our preference for giving equal consideration to each basis trajectory.
We find the optimal set of blend function values and derivatives as follows,

\begin{equation*}
\begin{aligned}
& \underset{\mathbf{W}}{\text{minimize}}
& & \sum_{k=0}^n \Big( (w_k - \frac{1}{2})^2du \; + \; \lambda( \frac{d^4 w}{du^4})^2du \Big) &&\\
& \text{subject to}
& & \mathbf{w}_{k+1} = \mathbf{w}_{k} + (\mathbf{M}\mathbf{w}_{k} + \mathbf{N}\mathbf{v}_{k})du &&\\
\end{aligned}
\end{equation*}
\begin{equation} \label{eqn:optimization}
\begin{aligned}
0 \leq w_k & \leq 1 \\
v^{\text{min}} \leq v_k & \leq v^{\text{max}} \\
\|\vec{\sigma}_k - \vec{P}_0\| & \geq d_{min}\\
\|\vec{\sigma}_k - \vec{P}_1\| & \geq d_{min}\\
\end{aligned}
\end{equation}
\begin{equation*}
\begin{aligned}
& \text{where} & & \vec{\sigma}_k = w_k\cdot\vec{\sigma}_A(\frac{k}{n}) + (1 - w_k)\cdot\vec{\sigma}_B(\frac{k}{n})  \\
\end{aligned}
\end{equation*}

The problem in (\ref{eqn:optimization}) is nonconvex,
and is therefore sensitive to initialization.
We initialize our solver with a default initial blend 
that averages the input trajectories exactly,
with weights set to $\frac{1}{2}$.

\paragraph{Performance} In our implementation,
we solve the problem in (\ref{eqn:optimization}) 
using the commercially available non-convex solver SNOPT~\cite{gill:2002}.
We rely on SNOPT to numerically calculate 
the Jacobian matrices of our optimization problem.  In all our experiments, we discretize $w$ at a moderate resolution of $n=50$ samples.
We experimentally find that
we can solve this optimization problem in under 500ms 
on a 2.8 GHz Intel Core i7 processor for all our shots.

\subsection{Generating the Final Trajectory}

Our final path is parameterized in 
terms of the path parameter $u$, and not time.
We apply an easing curve to our path to generate a smooth final trajectory,
using the method shown in Joubert et al.~\shortcite{Joubert15}.

So far we have only computed the look-from trajectory.
We still need to generate the look-at and field of view trajectory. 
To do this, we linearly interpolate 
between the two start and end look-at points and field of view values,
and then apply the same easing curve.


Once we have a final camera trajectory,
we check it against our quadrotor model for violations of physical limits using our open-source Flashlight library~\cite{flashlight:2016}.
If any exist,
we linearly time-stretch the easing curve until no constraints are violated.
Practically speaking,
we conservatively set the total time of our transitions 
to avoid violating physical constraints.
We have found that our default easing curve 
rarely produces trajectories that exceed these limits.
\section{Tracking and Control Platform} \label{sec_tracking}


\begin{figure}
\includegraphics[width=3.3in]{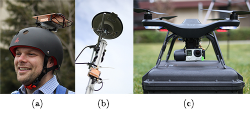}
\caption{Overview of the major physical components of our hardware platform. A subject (a) wears a position and orientation tracker on a helmet. GPS corrections are provided from a base station (b). The quadrotor (c) is equipped with an orientable camera and similar tracking hardware. 
\label{fig:hardware}}
\end{figure}

\begin{figure}[tb]
\centering
\includegraphics[width=1\linewidth]{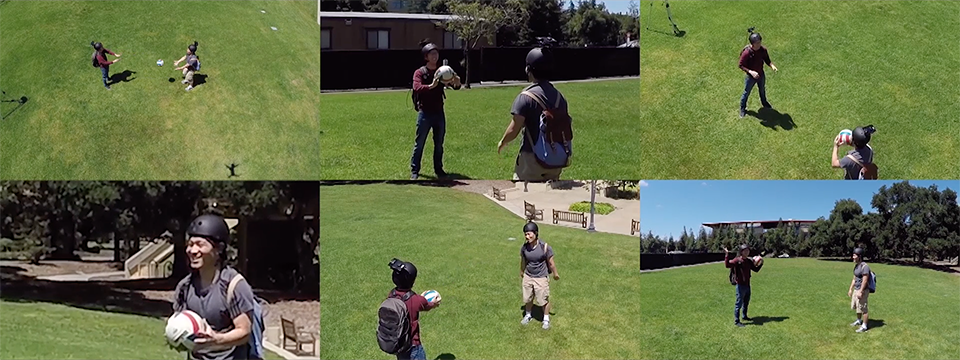}
\caption{Here we show a sequence of static shots captured by our system during a single shoot of two subjects playing catch. Top left to bottom right: apex from above, external of subject in red, external from above of subject in red, internal of subject in gray, external of subject in gray, apex. Notice the line of action is maintained throughout these shots: The person in red is always on the left, and the person in gray remains on the right.
\label{fig:staticshotsequence}}
\end{figure}

\begin{figure*}[th!]
\centering
\includegraphics[width=1\linewidth]{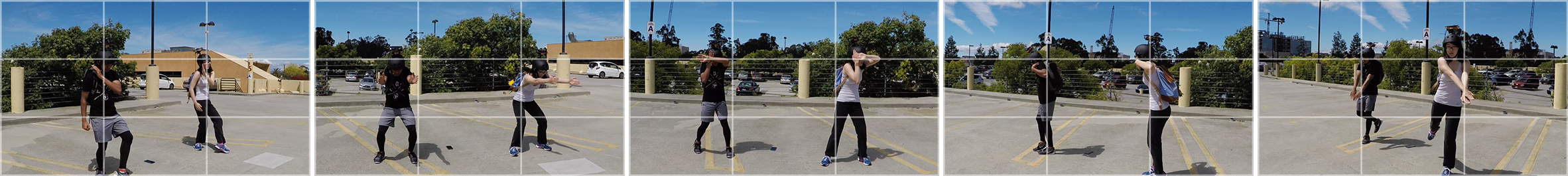}
\caption{Here we show a sequence of frames from a transition captured by our system while filming a choreographed dance routine. This transition goes from an external shot of the right character to an external shot of the left character.
\label{fig:transitionframes}}
\end{figure*}

Our system has to maintain a virtual representation of the scene to plan shots, and place a quadrotor camera accurately according to this plan. Here, we present a hardware platform that achieves these goals (Figure~\ref{fig:hardware}). We place active trackers containing an RTK GPS and IMU module on each subject and the quadrotor. The RTK GPS module uses a stream of corrections from a GPS base station to produce a centimeter-accurate position estimate. Additionally, the IMU module estimates the orientation of the tracker. Specifically, both the IMU and GPS output is fed into a Kalman filter to estimate the current position and orientation state. This state is communicated to a central computer using a low latency network. The central computer executes our shot planning algorithm, producing a quadrotor trajectory. The trajectory is turned into a sequence of control commands, sent to the quadrotor camera using the same wireless network.

RTK GPS modules are traditionally used in high-end surveying applications. We use one of the first affordable consumer-grade modules, the \textsc{Piksi RTK GPS} from Swift Navigation~\shortcite{swiftpiksi:2013}. An RTK GPS achieves single centimeter position accuracy by analyzing the carrier wave of the received GPS signal, and comparing it to the same signal received by a base station. This technique is quite sensitive to occlusions of the satellite constellation. Because of this sensitivity, we use our system in large open environments.

We use off-the-shelf long range radios to communicate between the various components of our platform. Specifically, we use the \textsc{Ubiquiti Bullet M5} radio. This radio provides sub 5 ms communication latency over a range of several hundred meters, enabling our system to maintain an up-to-date virtual representation of the scene.

The quadrotor we use in this paper is a modified \textsc{3DR Solo}, carrying a gimbal-mounted \textsc{GoPro Hero 4 Black} camera, and a RTK GPS module. This quadrotor uses the \textsc{APM} autopilot software~\cite{apm:2015}, which includes an onboard Kalman filter for position estimation. We extend and tune the Kalman filter to accept position estimates from an RTK GPS module. To fly a quadrotor camera according to a look-from and look-at trajectory, we use the same trajectory follower as Joubert et al.~\shortcite{Joubert15} to drive the onboard control system. High-accuracy position estimates from the RTK GPS aid the control system in placing the quadrotor accurately.

In our results, we present experimental evidence to support the efficacy of this platform for autonomously capturing cinematography.





\section{Results} \label{sec_results}

In order to test our system, we captured footage using our system of a range of scenarios including taking a selfie, playing catch, and performing a choreographed dance routine. We show several shots produced by our system in Figure~\ref{fig:teaser}. We encourage readers to also view the paper video, which walks through the results in detail. 

\paragraph {Well-composed Static Shots}
We present examples of several static shots generated using our system in Figure~\ref{fig:staticshotsequence}.
These shots are captured from a single flight, and feature two subjects playing catch.
Each of these shots respect the rule of thirds and our safety constraints, and are cropped to match the intended compositions.
Furthermore, our system maintains the line of action over successive shots. As a consequence, in each of these shots, the relative left-right positioning of subjects in frame is consistent.


\paragraph {Safe and Visually Pleasing Transitions}
Our transition planner is able to produce transitions that are both safe and visually pleasing.
Figure~\ref{fig:transitionframes} shows a set of still frames from a transition captured using our system.
Both subjects change smoothly in size, and move reasonably in screen space through the transition.
See the video for more examples to best qualitatively evaluate the transitions produced using our system. 
Our video includes transitions where there is a more significant change in crop.


\paragraph {Capturing Scripted Scenarios}
We also used our system to capture a fully scripted scenario. We staged a simulated graduation ceremony, and captured multiple takes of the entire performance, repositioning the camera between takes. Before a take, we pose our actors for a specific visual framing and autonomously place the camera. An editor used this footage to create a short narrative, cutting between different angles as the action smoothly unfolds. This use case demonstrates how our system can be used as part of the traditional cinematography process. 


\paragraph {Imposing Safety Constraints}

\begin{figure}[b!]
\centering
\includegraphics[width=3.3in]{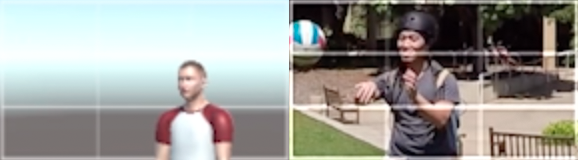}
\caption{A failure case, where the suggested crop does not match the target framing. $\alpha = 14.9$ while $\alpha_{\text{max}} = 50$.
\label{fig:badalignment}}
\end{figure}

Our decision to prioritize safety causes some failure cases where we do not manage to frame a subject accurately. These failure cases occur when we attempt to capture a close shot from far away. These situations are particularly challenging, since small errors in orientation can significantly impact the visual composition. Internal shots are particularly sensitive to this effect. If the primary subject is looking at another subject, the internal has to be placed behind the other subject to capture the face of the primary subject. Figure~\ref{fig:badalignment} shows an example of this occurrance. The internal shot is cropped significantly, and the resulting footage does not manage to respect the rule of thirds.

\paragraph {Accurate Tracking and Control}

\begin{table}
\centering
\caption{
Position error of RTK GPS compared to a conventional GPS fused with an IMU and barometer. We separately report altitude noise compared to the barometer. We ran two experiments, each consisting of five 5-minute trials. First, we held both trackers stationary, and measured position noise. Then we performed a loop closure test by moving the tracker in a random pattern before bringing the tracker back to the starting point. RTK GPS outperforms the conventional tracker by one to two orders of magnitude throughout. CEP\textsubscript{95} is defined as the radius of a circle within which 95\% of samples fall.\label{table:gpsprecision}}
\begin{tabular}{| c || c | c |}
\hline
 & Ours & Conventional \\
\hline \hline
North-East CEP\textsubscript{95} & 0.017 m & 1.68 m \\
\hline
Altitude Std. Dev. & 0.020 m & 0.108 m \\
\hline
Distance Error after Loop Closure & 0.011 m & 1.058 m \\
\hline
\end{tabular}
\end{table}

We performed a series of experiments on our tracking system, comparing it against a tracker from a consumer quadrotor, consisting of a conventional GPS fused with an IMU and barometer. We report results in Table \ref{table:gpsprecision}. In each of these experiments, the RTK GPS outperformed the conventional tracker by one to two orders of magnitude in position. 

We also test the hover accuracy of a quadrotor using our tracking system. We use RTK GPS as the ground truth to measure drift of the quadrotor under the control of either our tracker or the conventional system. The two orders of magnitude more accurate results of RTK GPS compared to standard GPS (Table \ref{table:gpsprecision}) motivate our decision to use RTK GPS as ground truth in this experiment. Using RTK GPS, the quadrotor remained within 0.35m for 95\% of the flight time. Using conventional GPS, it remained within 1.05m.



\begin{figure}[thb!]
\centering
\includegraphics[width=1\linewidth]{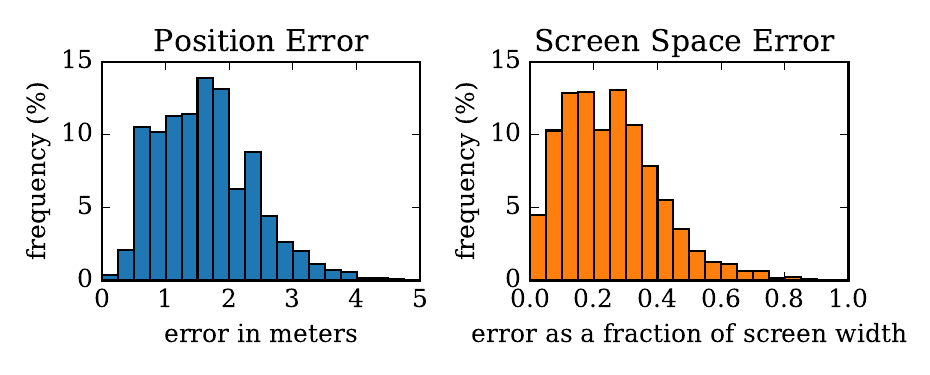}
\caption{World space and screen space error incurred when using conventional GPS to track subjects and plan shots. We track subjects through an 8 minute session using both RTK and conventional GPS. We use RTK GPS as ground truth, and conventional GPS to plan shots. Conventional GPS produced world space error of several meters, potentially violating our safety constraint. We automatically planned a virtual camera shot every 4 seconds, and report the resulting screen space error. Using conventional GPS incurs unacceptable screen space error, potentially placing the subject halfway across the frame or more.
 \label{fig:positionandscreengpserror}}
\end{figure}

We also investigate the screen space impact of conventional versus RTK GPS, reported in Figure~\ref{fig:positionandscreengpserror}.
 Significant screen space error is incurred when using conventional GPS to capture our set of shots, validating our design choice to use a higher accuracy approach.

\section{Discussion and Future Work} \label{sec_discussion}

In our work, we consider the placement and size of subjects in screen space. However, we plan paths in world space, only indirectly controlling the screen space behavior of subjects.
An exciting follow-up to our transition planner is an algorithm for directly controlling screen space behavior of subjects, solving for the equivalent camera path.


Our tracking system in its current form is fairly bulky and intrusive. Our trackers are a prototype, and the assembly can easily be miniaturized. RTK GPS is also under active development, with modules becoming more robust and affordable, and base stations being offered as a cloud service. Further, we designed our tracking system to operate independently of the camera. An exciting path forward is to additionally control the quadrotor by visually tracking the primary subject whenever she is in frame. 





We made a set of simplifying assumptions for the scenarios we considered. That is, we limited ourselves to up to two subjects, and assumed the subject stays within a fixed safety sphere during filming. The next step towards a drone cinematographer is to lift both these restrictions.

Currently our system does not attempt to aggressively follow or respond to people's movement. We are interested in extending our system to capture moving versions of our static shots while maintaining the safety of our subjects. Given the tight framing of our shots, we imagine that doing so is a nontrivial problem, potentially addressed using concepts from model predictive control~\cite{tedrake:2014}.

In this paper we considered the composition of shots. There are also many other factors that play into producing aesthetic footage, such as lighting and color. Broadly speaking, we think that incorporating aesthetic considerations into quadrotor camera control can significantly alter the way people produce video.

\section{Conclusion} \label{sec_conclusion}

We presented a system that attempts to follow composition principles when autonomously capturing footage of people with a quadrotor.
Along the way, we have encoded a set of canonical shots from cinematography literature and adapted them to respect safety constraints.
We also presented a novel transition planner that produces visually pleasing transitions while also satisfying our safety constraints, and quadrotor dynamics.
Our implementation is built on a tracking system that uses RTK GPS to localize the positions of both subjects and the quadrotor camera with centimeter accuracy. Finally, we successfully captured multiple scenarios with reasonable accuracy using a real quadrotor camera.

\section{Acknowledgements} \label{sec_acknowledgements}

We would like to thank the Siggraph Asia reviewers for their constructive comments. Thank you to Mira Dontcheva, David Salesin, and Fergus Noble for the helpful discussions about the technical components of this work. We also thank all the actors and dancers featured in our results. Thanks to Matthew Cong for valuable technical writing assistance and enthusiasm for our work. This work was made possible by the generous donations and support of both 3D Robotics and Swift Navigation, as well as a Microsoft Graduate Women's Scholarship.


\bibliographystyle{acmsiggraph}
\bibliography{000_main}

\end{document}